\documentclass[conference, 10pt]{IEEEtran}

\usepackage{amsmath,amsfonts}
\usepackage{algorithmic}
\usepackage{algorithm}
\usepackage{graphicx}
\usepackage{textcomp}
\usepackage{threeparttable}
\usepackage{xcolor}
\usepackage{multirow}
\usepackage{pifont}

\begin{document}
	
	\bstctlcite{IEEEexample:BSTcontrol}
	
	\title{Empowering Malware Detection Efficiency within Processing-in-Memory Architecture}
	
	%\title{Processing-in-Memory Architecture with Precision-Scaling for Malware Detection}
	\vspace{-2em}
	
	\author{\IEEEauthorblockN{Sreenitha Kasarapu$^*$, Sathwika Bavikadi$^*$, Sai Manoj Pudukotai Dinakarrao}
		\IEEEauthorblockA{Department of Electrical and Computer Engineering, George Mason University, Fairfax, VA, USA\\
			\{skasarap, sbavikad, spudukot\}@gmu.edu}}
	
	\maketitle
	
	\begingroup\renewcommand\thefootnote{\textsection}
	\footnotetext{*Both authors contributed equally to this research}
	\endgroup

	\begin{abstract}
		
		% The wide adaptations of embedded systems in multiple fields have led to smart connectivity across devices and enhanced computation capabilities. Despite the vast applications in different areas embedded systems tend to face huge security threats. One of the critical security vulnerabilities is caused by malicious software \textit{a.k.a} malware. Malware detection techniques based on Machine Learning are becoming popular in recent times. As DNNs and CNNs are quite efficient in image processing. The only downside of neural network (NN) architectures is their massive need for computational resources. To keep the malware detection model updated they need to be trained with newer malware and benign samples frequently. However, the frequent need to train a model with immense resource consumption for real-world applications is a strenuous task. To address such concerns, we introduce a PIM-based architecture to improve memory access latency. This helps restrain the number of resources consumed each time the malware detection model is updated. To further improve the throughput and energy consumption we employ precision scaling for CNN models. Our proposed PIM architecture has 1.09$\times$ higher throughput than other LUT-based PIM architecture. Furthermore, precision scaling and PIM improve the energy efficiency by 1.5$\times$ compared to the full-precision operation without any penalty in performance.  

		The widespread integration of embedded systems across various industries has facilitated seamless connectivity among devices and bolstered computational capabilities. Despite their extensive applications, embedded systems encounter significant security threats, with one of the most critical vulnerabilities being malicious software, commonly known as malware. In recent times, malware detection techniques leveraging Machine Learning have gained popularity. Deep Neural Networks (DNNs) and Convolutional Neural Networks (CNNs) have proven particularly efficient in image processing tasks. However, one major drawback of neural network architectures is their substantial computational resource requirements. Continuous training of malware detection models with updated malware and benign samples demands immense computational resources, presenting a challenge for real-world applications. In response to these concerns, we propose a Processing-in-Memory (PIM)-based architecture to mitigate memory access latency, thereby reducing the resources consumed during model updates. To further enhance throughput and minimize energy consumption, we incorporate precision scaling techniques tailored for CNN models. Our proposed PIM architecture exhibits a 1.09$\times$ higher throughput compared to existing Lookup Table (LUT)-based PIM architectures. Additionally, precision scaling combined with PIM enhances energy efficiency by 1.5$\times$ compared to full-precision operations, without sacrificing performance. This innovative approach offers a promising solution to the resource-intensive nature of malware detection model updates, paving the way for more efficient and sustainable cybersecurity practices.

	\end{abstract}

%\vspace{-1em}
\section{Introduction}

With the technical developments in hardware architecture and embedded systems, IoT applications have procured enormous interest in the past few decades \cite{iot-1}. The immense desire to automate user applications and interactive software systems such as smart homes, smart grids, and digital monitoring has led manufacturers to produce massively. These automated devices connect to the internet over a network for communicating between devices. As these devices have medical and activity-tracking abilities, it is important for them to communicate. They handle vast amounts of user data daily and are targeted by cyber-attackers. These systems are vulnerable to security threats \cite{Abbas2016BigDI} due to malware. Malware is malicious software developed to infect a system to explore and steal information such as passwords, and financial data, and manipulate the stored data without the user's consent. In the year 2021 alone, there were more than 5.4 billion recorded malware attacks \cite{stat_1}. The first half of the year 2022 had 2.8 billion malware attacks.

Despite the advanced anti-malware software, malware attacks increase in millions each year \cite{stat_1}. This is due to the newer emerging malware each year. Adversaries generate millions of new signatures of malware each year \cite{stat_2} to steal valuable information for financial benefit and stay undetectable. The massive increase in cyber attacks due to malware poses a huge threat to hardware security \cite{cyber-risk-2}. The exploitation of confidential user information leads to substandard user experience, so it is vital to detect the malware. Realizing the threat caused by malware in terms of access to sensitive information, stolen information, and billions of revenue loss, severe measures are being taken to abate malware escalation.

Static and dynamic analysis \cite{sta_dy} is employed for malware detection. Static analysis \cite{sta_dy} is performed in a non-runtime environment by examining the internal structure of malware binaries and not by actually executing the binary executable files.  
In dynamic analysis, the binary applications are inspected for malware traces by executing them in a harmless, isolated environment \cite{sta_dy}. Unlike static analysis, dynamic analysis is a functionality test. The static analysis serves as quick testing but is not efficient. Though efficient dynamic analysis is a bulky and time-consuming process.

Malware detection using Machine Learning (ML) is seen as an efficient technique \cite{img_vis}. A variety of Machine Learning (ML) and Federated Learning techniques \cite{sanket_fl} depict superior malware detection capability than the static and dynamic analysis methods \cite{nataraj}.  
Among the ML-based malware detection techniques, the CNN-based image classification technique \cite{cnn_detect} is more robust and efficient due to its prime ability to learn image features.
However, one of the main challenges with adopting such a technique is the requirement of enough samples for training. 
With the exponential increase in the generation of newer malware families each year, it is complex to train a model on heterogeneous data and update it frequently. With the increase in the need for updating, the training overhead increases consistently.

In this study, we propose a novel approach for malware detection to utilize the in-memory computing technique.
Processing-in-memory (PIM) is a novel computing paradigm in which the memory chip is enhanced with computing capabilities. This essentially restricts the circulation of the data within the memory chip and thereby drastically minimizes the power consumption and latency caused by the data movements. In addition, a PIM architecture takes advantage of its proximity to the data to perform massively parallel computing. Therefore, such a PIM paradigm is particularly suitable for data-intensive applications like deep learning (DL) and optimization problem.

Several recent studies have proven that PIM architectures outperform GPU and CPU designs for training deep neural networks (DNN) and combinational optimization problems in terms of throughput and energy efficiency. While traditional PIM architecture, like bitline-wise architecture \cite{Ambit} and analog crossbar array architecture \cite{mramdima}, 
have been regarded as better alternatives to conventional computing hardware for executing the heavy computational load of DNN.
These architectures suffer from the complexity and overhead associated with digital-to-analog (DAC) and analog-to-digital (ADC) conversions. Unlike the bitwise processing PIMs, the recently developed Look-up-table based (LUT-based) PIMs have been found to be more flexible, with superior energy efficiency for a similar level of performance, such as LAcc \cite{LACC} and pPIM \cite{pPIM}. This feature makes the LUT-based PIM architecture more suitable for performing adversarial attack generation.

In this work, we address all the issues mentioned above. We propose a technique that can effectively detect malware with limited resources. We propose a PIM-based technique that changes the memory access capabilities. Thus improving the inference throughput. We further employ precision scaling to decrease power consumption.

The novel contributions of this work can be outlined in a three-fold manner: 
\vspace{0.0003em}
\begin{itemize}

	\item A memory-efficient malware detection by using an in-memory computation technique.
	
	\item Precision scaling to decrease the power consumption of malware detection model.
	\item Scaling malware samples to 16-bit, 8-bit, and 4-bit integer types and still retaining a malware detection accuracy of 98\%.

\end{itemize}

The rest of the paper is organized as follows: Section \ref{related_work} describes the related work and its shortcomings and comparison with the proposed model. Section \ref{problem} describes the problem formulation. Section \ref{prop_tech} describes the proposed architecture, which assists with the proposed PIM-based malware detection model training. The experimental evaluation of the proposed model and comparison with various ML architectures is illustrated in Section \ref{exp_results}, and then we conclude in Section \ref{conclusion}.

%\vspace{-0.7em}

\section{Related Work}
\label{related_work}
\subsection{Malware Detection Techniques}

%Traditionally malware detection is carried out using static and dynamic analysis. Static analysis \cite{sta_dy} is performed in a non-runtime environment by examining the internal structure of malware binaries and not by actually executing the binary executable files. Static analysis is not an efficient approach for malware detection but serves as a quick testing tool \cite{static_limits}. In dynamic analysis, the binary applications are inspected as malware, or benign by executing in a harmless, isolated environment \cite{sta_dy,dynamic}. Unlike static analysis, dynamic analysis is a functionality test. Although dynamic analysis can handle complex malware, it can only expose malware that exhibits malware properties \cite{dynamic}. Dynamic analysis is not efficient in detecting hidden malware code blocks which restrain it from getting executed, and it is a time-consuming process.
Static analysis \cite{sta_dy} on malware data is performed by comparing the opcode sequences of binary executable files, control flow graphs, and code patterns. The main drawback of static analysis is, it cant detect malware when adversaries add junk of unrelated functionalities, which decreases the malware similarity score \cite{static_limits}.
Malware detection using dynamic analysis is performed based on detecting system calls or HPC \cite{sta_dy}. But they are not efficient in detecting hidden malware code blocks and are computationally expensive.

Later \cite{nataraj} introduced a technique for malware detection using image processing where binary applications are converted into grayscale images. The generated images have identical patterns because of the executable file structural distributions. The paper used the K-Nearest Neighbour ML algorithm for the classification of malware images. Other approaches \cite{img_vis} include image visualization and classification using machine learning algorithms such as SVM. However, these approaches don't address the problem of classifying newer complex malware that is code obfuscated, polymorphed, etc. Neural networks such as ANNs are used extensively to solve the problem \cite{img_process}, as neurons can capture the features of the images more accurately than other machine learning algorithms. But, the fully connected layers of artificial neural networks tend to exhaust computational resources. 
In \cite{cnn_detect, sanket_abhijitt_date2021, sanket_cases_2019, sanket_dac_2021, sanket_date_2023, sanket_dhavlle2021novel, sanket_glsvlsi_2022, sanket_iccd_2022, sanket_icmla_2019, sanket_ictai_2019, sanket_isqed_2024, sanket_rram_glsvlsi_21, Sanket_DATE'20, sreenitha_kasarapu2021demography, sreenitha_mdpi, sreenitha_sanket_aspdac, sreenitha_sanket_glsvlsi, sreenitha_sanket_tcad, sreenitha_sanket_ubol, sreenitha_sathwika_vlsid, raghul_iscas, raghul_RS2020, raghul_SaravananICDSMLA'19, raghul_SaravananICICNIS'21, raghul_trng2020, Raghul2019} authors used Convolutional neural networks, as they are popular for their ability to efficiently handle image data through feature extraction by Convolutional 2D layers and using Maxpooling 2D layers to downsample the input parameters, thus, reducing the computational resources. The drawback here is, they need to be trained with a balanced dataset to perform classification efficiently, but with an increase in malware families, collecting each type of malware for training is challenging.

\subsection{Processing-in-Memory (PIM)}

In recent years, PIM designs have received a lot of attention from DNN/CNN applications. PIMs have been regarded as a better alternative to conventional computing hardware for executing the heavy computational load imposed by the convolutional layers of a CNN. In contrast to the CPU or GPU-based architecture, computational functions for the PIM architecture are executed in the memory itself. Thus, this eliminated the need for time- and energy-intensive data movement. Consequently, the PIM architecture can reduce the latency and energy costs associated with data movement. 

Moreover, with the integration of memory and processing capability, the PIM architecture is able to efficiently execute matrix-vector multiplication (MVM) operations, which are fundamental computing operations in various disciplines of research such as signal processing, machine learning \cite{pPIM}, deep learning \cite{Tpds_pPIM, sathwika_LUT, sathwika_MLacc, sathwika_upim, Sathwika_ISQED}, and stochastic computing, image processing, and recognition \cite{upim}, data mining \cite{pPIM}, and cryptographic \cite{aes}.

Numerous works have been proposed on in-memory computing hardware accelerators on different memory platforms including the traditional memory platforms of 
Static and Dynamic Random Access Memory (SRAM \& DRAM) \cite{DRACC} as well as novel non-volatile Resistive RAM (ReRAM) \cite{PRIME}, Phase-changing Memory (PCM), and Magnetic RAMs such as Spin Transfer Torque MRAM (STT-MRAM) \cite{mramdima}, and Spin-Orbit Torque MRAM (SOT-MRAM) \cite{IMCE} technologies. It has been found that a satisfactory level of accuracy can be retained even despite performing various levels of quantization/down-scaling of data parameters in CNN algorithms \cite{Tpds_pPIM}. 
This opens up an exploration space for high-performance and low-power CNN implementations for real-time application domains such as IoT, mobile, and edge applications. 
The PIM architecture is gaining popularity in real-time application domains. To the best of the authors' knowledge, the PIM architecture has not been utilized in the context of malware detection.

%We, therefore, need an efficient model which can address all the concerns mentioned above. Also, the model must efficiently classify code obfuscated and stealthy malware without the need for a vast training dataset. In this paper, we address this by developing a code-aware generator that can generate realistic images. These generated images can replicate features of various malware families, solving the problem of training the CNN for efficient malware detection with limited samples.  

\section{Problem Formulation}
\label{problem}
\vspace{-0.25em}

With technology advancements, attackers are introducing, any complex malware families, making it impossible for embedded systems to keep track. Even advanced anti-malware software fails to detect these advanced malware families \cite{morphism}. One can define the problem of reliable malware detection in embedded devices as follows:

\vspace{0.001 em}
\begin{equation}
	\centering
	\mathbb{D} \leftarrow [{B}^m_1, {M}^n_1, {B}^m_2, {M}^n_2, {B}^m_3, {M}^n_3, \cdots, {B}^m_q, {M}^n_r]
	\vspace{0.0001em}
	\label{eq1}
\end{equation}

As shown in Equation \eqref{eq1}, B represents benign and M represents the malware executables for embedded systems. The $B^m$ represents the $m^{th}$ benign sample, ${B}^m_q$ represent the $q$ number of patterns in a given $m^{th}$ sample. Similarly, $M^n$ represents the $n^{th}$ malware sample, ${M}^n_r$ represent the $r$ number of patterns in a given $n^{th}$ sample. These benign and malware are stored in dataset $\mathbb{D}$.

\vspace{0.001 em}
%\vspace{-1em}
\begin{equation}
	\begin{aligned}
		&& C(\mathbb{D}): {X} \rightarrow {Y} \hskip 0.75em \\
		& \text{s.t.} 
		&\mathbb{D} \leftarrow  \iint_{a}^{b} (B^m_q, M^n_r) \,dq \,dr
		%   \vspace{0.0001em}
		\label{eq2}
	\end{aligned}
\end{equation}

%\textcolor{red}{Problem formulation is incomplete. For instance, what if the constraints in (2) are not satisfied?}
In Equation \ref{eq2}, $C$ is a classifier model trained with dataset $\mathbb{D}$ to perform malware detection. 
%The dataset $\mathbb{D}$ contains, a combination of malware and benign  samples.
After training, the classifier $C$ will be able to classify any input sample $X$ and map it to either malware 
class ${M}$ or benign class $B$. The output class is represented as $Y$. New benign and malware samples are needed to train and keep the malware detection model updated. As represented by the integral of $(B,M)$, for each new pattern of benign $q$ and malware $r$, the classifier $C$ needs to be updated. The limit $a$, $b$ represent the newer patterns of data to be added for training. 
%\textcolor{red}{What are m and n?}
%But, the memory required to perform inference, represented as $\mathfrak{Mem}[\mathbb{C}]$ should be less than the available resources in an IoT node, represented as $\mathfrak{Mem}[node]$. \revision{If the constraint in equation \eqref{eq2} is not met, then the inference task cannot be carried out by the device.
	
	\vspace{0.001 em}
	\begin{equation}
		\centering
		S = f_0 + \sum_{i=1}^{d} f_1\cdot P^{-i} + \sum_{j=1}^{d} f_2\cdot I^{-j}
		\vspace{0.0001em}
		\label{eq3}
	\end{equation}
	
	So the problem that arises with consistent data updation i.e,. the need for huge computational resources can be formulated as shown in Equation \ref{eq3}. We build a dual optimization problem, by considering input data precision $P$ and in-memory data processing steps $I$. The aim is to optimize these both parameters, as by doing so we can decrease the computational resources.  Where, $f_0$, $f_1$ and $f_2$ are optimization constants. The $d$ represent the size of the dataset $\mathbb{D}$. For each data point ranging from $(1,2, \cdots, d)$ the input precision $P$ and the in-memory processing speed $I$ is optimized. 
	
	Our proposed technique solves this by introducing a novel malware detection framework using PIM-based memory access, processing technique and also employing input precision scaling. These improve the resource consumption and throughput of machine learning model used for malware detection.
	%So the need for efficient malware detection with limited resources arises.
	%\textcolor{red}{If I were you, I would put the second part of equation as constraint (s.t.)}

\section{Proposed Technique}
\label{prop_tech}

\vspace{0.0001em}

\begin{figure*}[ht]
	
	\includegraphics[width=\textwidth, height = 5.5cm]{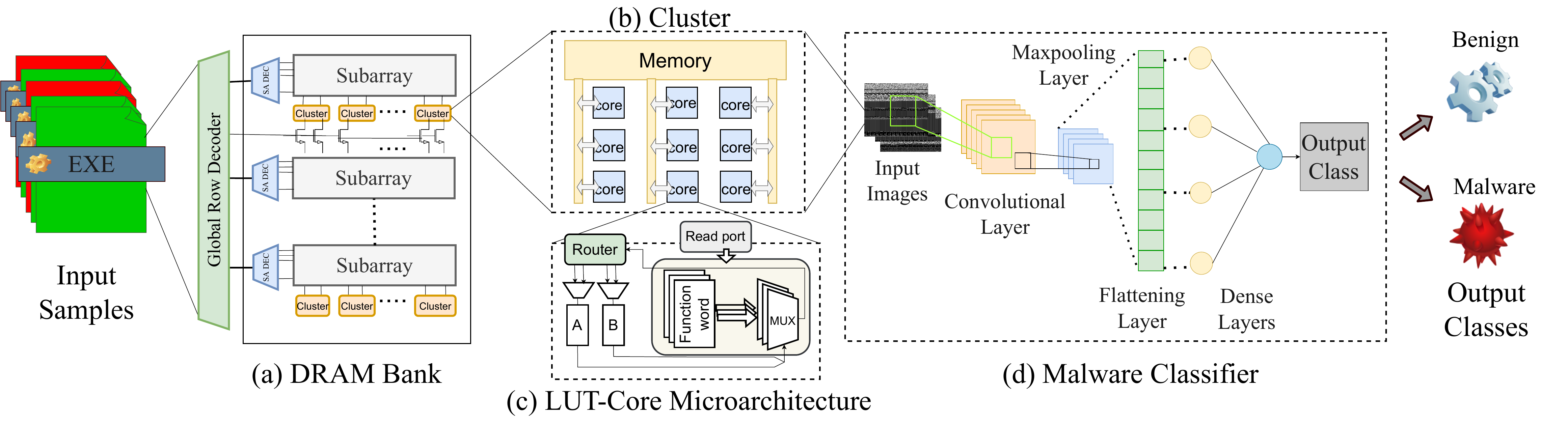}
	
	\caption{Hierarchical view of the architecture implementation of malware detection on the processing in-memory architecture }
	
	\label{fig:overview}
\end{figure*}

\subsection{Overview of the Proposed Technique}\label{AA}

The overview of the proposed technique follows the flow as shown in Figure \ref{fig:overview}. The input data samples are stored in the DRAM memory bank represented in Figure \ref{fig:overview}(a). In-memory processing is employed using a DRAM cluster to improve memory access time. The architecture of each DRAM cluster is represented in \ref{fig:overview}(b). Each cluster has the LUT-core represented as \ref{fig:overview}(c). The input binary samples are processed in the memory and converted into images. As this is done using an in-memory processing technique, there was no need for excessive data movement. Once the data is accessed from memory it is given as input to the malware classifier given in \ref{fig:overview}(d). The test data used for inference is precision scaled using uniform quantization. The different elements in the proposed technique are:

\vspace{0.2em}
\begin{itemize}
	% \item \emph{Data Updation}: Variations patterns of benign and malware samples are collected for training. To make the malware detection model efficient it is automatically updated using heterogeneous data.

	\item \emph{PIM Unit}: A novel PIM architecture is employed to decrease the memory access latency. In-memory processing would improve throughput and limit resource consumption.
	
	\item \emph{Precision-scaling}: While retaining the malware detection accuracy, a low-precision version of data is employed to decrease the power consumption.
	
	\item \emph{Malware Classifier}: Optimized input samples are fed to the Convolutional Neural Network for training the classification model.

\end{itemize}

\subsection{Malware Detection Model}

Computer vision-based machine learning (ML) techniques need images for localized feature extraction. So the input binaries (malware and benign) are converted into grayscale images. The binary applications files are converted into a raw binary bitstream. This binary bitstream is then converted to an 8-bit vector. Each 8-bit vector containing the binary values is taken as a byte, representing different image pixels. Each of these 8-bit vectors is arranged to form an image. The size of the gray-scale image varies with the size of the binary file. To train a CNN, the input data must be uniform. To address this challenge, we perform image resizing and scaling in order to make its size uniform. As
the pattern or sub-pattern of malware cannot alter despite
embedding the malware to launch malicious payload, through
this technique the malware can be detected with a higher
performance (around 98\% accuracy).

%In the computer vision-based detection technique, the application binary is converted into a grayscale image for localized feature extraction. A raster scanning is performed on the converted binary images to find the imagpatterns. Each pattern is of 32×32 block size. We utilizea cosine similarity to distinguish between multiple patternsi.e., if the cosine similarity of two patterns is higher thanthreshold (0.75 in this work based on conducted experiments),they are considered to be same. When more than one matchedpatterns are found, the one with the highest cosine similarityis considered. It needs to be noted that the pattern matchingfor an incoming binary is performed with the patterns inthe database (created through similar process during trainingphase, but offline). Once the image patterns are recognized fora given binary file, the whole image binary is converted into asequence of patterns (Each pattern is provided with a uniqueID). This sequence is fed as input to the convolutional neural network (CNN). 

Using this process, a training dataset is generated as shown in equation (1) which consists of several sequences for a variety of
classes of malware (backdoor, rootkit, trojan, virus, and worm)
and benign applications. So, a CNN model is trained using these data samples for classification as shown in Figure \ref{fig:overview}(d). As the processing is done in memory, this data is not circulated within the memory chip. Thus, decreasing the computation steps and resources. Feature extraction is done on the images using the convolutional 2D layers of the CNN architecture, followed by the max-pooling 2D layers. The data is then flattened to pass it to the dense layers, to decide the output class. The mac operations in each of these layers are accommodated by PIM architecture by programming the LUT cores inside the cluster as shown in Figure \ref{fig:overview}(c). The CNN classifier trained on input data can be defined as shown in equation \ref{eq2}. The classifier $\textbf{C:}$ trained on data $X$, maps the input to the class label $Y$.

The classification accuracy $a(\cdot)$, can be defined as the difference in probability of the predicted class $P(Y_{pred})$ to the real class $P(Y_{true})$.

\vspace{-0.5 em}
\begin{equation}
	\centering
	a(\cdot) \leftarrow P(Y_{true}) - P(Y_{pred})
	\vspace{0.1em}
	\label{eq4}
\end{equation}

\subsection{The PIM Architecture}

This work utilizes a PIM architecture designed to support compute-intensive applications including convolution neural networks (CNNs) and DNNs. This PIM architecture is depicted hierarchically in Figure \ref{fig:overview}, including (a) the arrangement of clusters within a DRAM bank, (b) the architecture of a cluster, and (c) the architecture of the LUT core. 

\subsubsection{Core Architecture}
In order to provide functional programmability, a LUT-based design is adopted for the PIM core instead of a pre-defined logic circuit. The LUT-based PIM is capable of performing in-memory arithmetic operations such as addition, multiplication, substitution, and comparison operations with a significantly lower delay compared to bitwise computations. A collection of these operations can therefore be used to implement various ML algorithms. 

Figure \ref{fig:overview}(c) shows a detailed view of the architecture of a single LUT core. The LUT cores inside the cluster are formed of an 8-bit 256:1 multiplexer, accompanied by eight 256-bit latch arrays. The pre-calculated outputs of any particular 8-bit operation are stored in the latches as eight 256-bit function words. These latches can read new function words from the bit lines of the DRAM subarrays. Each LUT can produce a 4-bit data output for two input data operands with a size of 4-bit width, as shown by A and B registers in Figure \ref{fig:overview}(c). These registers together drive the select pins of the multiplexers and make them ‘look up’ specific 8-bit data from the eight latches that represent the output of the operation.

\subsubsection{Cluster Architecture}

In order to perform operations necessary for CNN acceleration, such as convolution operations, the PIM cluster integrates all the operations done by the LUT core. Nine of the PIM cores are arranged as shown in Figure \ref{fig:overview}(b) and placed inside the memory unit to form a PIM cluster and perform in-memory computations. Inside the PIM cluster, the PIM core performs various logic and arithmetic operations associated with the CNN acceleration for malware detection. 
In order to perform a specific operation, such as multiplication and accumulation operations all the cores inside the cluster are connected by a router. The router also makes it possible to access data from any core at any moment throughout the implementation in order to perform the operations.
\subsubsection{Router Architecture}

All nine LUT cores in a cluster are connected via a router mechanism as shown in Figure \ref{fig:overview}(c), which enables direct and parallel communication between them. A router that connects the read/write ports on all components of the cluster in order to facilitate parallel communication among all the cores in the cluster. This enables the router to access any data at any point of implementation to perform operations required for CNN acceleration for malware detection. 

\subsection{Implementation on the PIM Architecture}

Malware detection performs similar mathematical operations as the CNN classifier, which contains max pooling layers and convolution layers with activation layers.
The PIM architecture can easily accommodate these operations by programming the LUT cores inside the cluster to perform these operations.
With the help of the cores and the routing mechanism, the PIM architecture can perform the mathematical operations required by the CNN layers.
Since the clusters in the PIM architecture are capable of performing the
required mathematical operations for CNN, an array of these clusters can be utilized to implement different layers of the CNN model.
The key advantage of using LUTs in a PIM architecture is that the LUT core in the PIM architecture can be re-programmed to perform an entirely different operation. As a result, the functional flexibility required for implementing malware detection using CNNs is provided.

\subsection{Data Format Optimization Using Precision Scaling}

Precision scaling is a process of scaling the precision point of input data to the desired decimal point. Precision scaling helps lower the number of mac operations, thus, decreasing the throughput and memory consumption. In this paper, we attain this by using uniform quantization. The input images (malware and benign) are stored in the DRAM memory bank as binary bits. As shown in Figure \ref{fig:overview}, while retrieving the data uniform quantization is applied. The test samples used for inference are quantized from floating point 32-bit to integer types 16-bit, 8-bit and 4-bit. This massively decreases the computations on top of the in-memory processing. The quantization process can be described through equation \ref{eq6}.

\vspace{0.001 em}
\begin{equation}
	\centering
	\mathbb{D}(r) = S\cdot (q-Z)
	\vspace{0.0001em}
	\label{eq6}
\end{equation}

In equation \ref{eq6}, q is the quantized value, S and Z are quantization parameters. Z is the zero point buffer, the same type as the quantized q. It may be used to pad zeros for the input quantized data so that it can meet the requirement of real value r. For 4-bit quantization, q is quantized as a 4-bit integer (for N-bit quantization, q is quantized as an N-bit integer). The quantized value q is written into the dataset $\mathbb{D}$ as a real number r. Each of the input images are quantized into either 16-bit, 8-bit or 4-bit integer types and written back into the dataset.

\vspace{-0.5em}
\section{Experimental Results}

\label{exp_results}

\subsection{Experimental Setup}

The proposed methodology is implemented on an Intel core Nvidia GeForce GTX 1650 GPU with 16GB RAM. We have obtained malware applications from VirusTotal \cite{rvirus} with 12500 malware samples that encompass 5 malware classes: backdoor, rootkit, trojan, virus, and worm. We collected about 13700 benign application files, which are harmless to work with. These binary files are converted into grayscale images of size 32 x 32 and fed to an ML model to build a malware detection classifier. We use the malware and benign samples to train a model such as AlexNet, ResNet18, ResNet34, ResNet50, VGG-16, and MobileNetV2. We analyze the malware detection accuracy, energy efficiency, and throughput of these models. Further, the inference accuracy of these models on precision scaling schemes such as 16-bit, 8-bit and 4-bit are evaluated.

%From malware families backdoor, trojan, virus, and worm, 12800 malware samples are generated. 

\subsection{Evaluation of Malware Detection Accuracy on Various Precision Schemes}

\begin{figure}[ht]
	\vspace{-10em}
	\includegraphics[width=0.5\textwidth, height= 11cm]{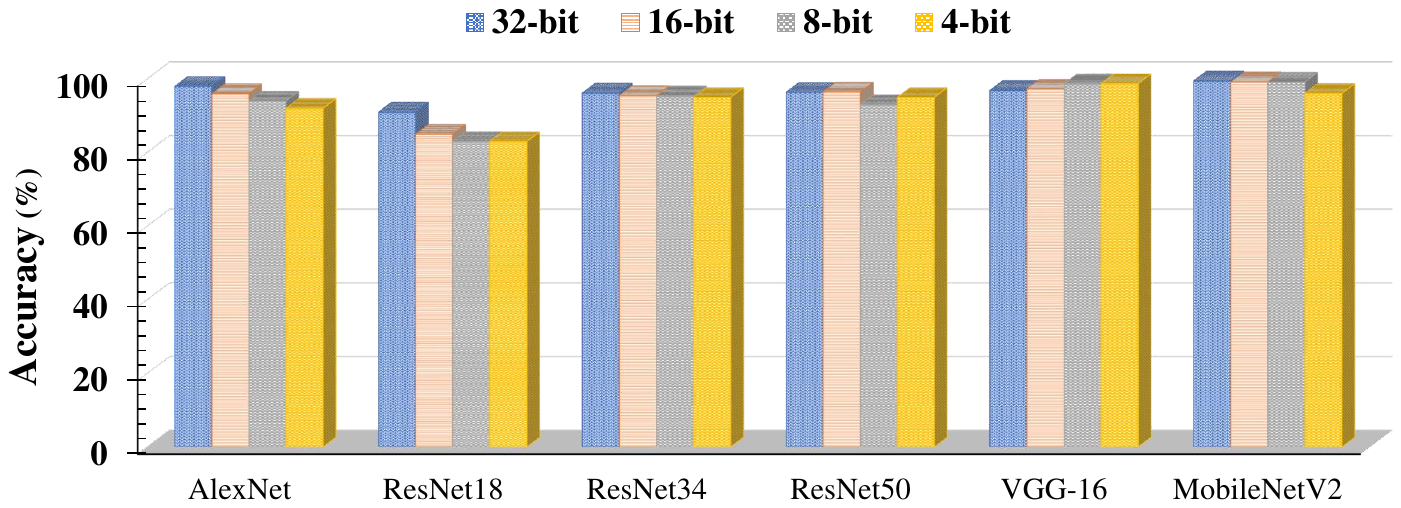}
	\vspace{-11em}
	\caption{Performance Evaluation of AlexNet, ResNet18, ResNet34, ResNet50, VGG16 and MobileNetV2 on the PIM accelerator with precision scaling (a) 32-bit floating point, (b) 16-bit integer type, (c) 8-bit integer type and (d) 4-bit integer type}
	\label{fig:accuracy}
	
\end{figure}

The integer precision schemes such as 16-bit, 8-bit and 4-bit are applied to data that is given as input to various ML algorithms. The effect of precision scaling on malware detection capability is observed in various algorithms. The base accuracy of models trained using 32-bit floating point data is compared to that of 16-bit, 8-bit and 4-bit data. As shown in Figure \ref{fig:accuracy}, the performance of various CNN models on different data precisions is compared. We can observe a considerable performance decay with the decrease in precision for models such as AlexNet and ResNet18. But other models such as ResNet34, ResNet50, VGG-16, and MobileNetV2 retain the accuracy despite the low precisions. We can observe a malware detection accuracy of about 98\% in models VGG-16 and MobileNetV2 for an 8-bit precision scheme. And for 4-bit precision models ResNet34, VGG-16, and MobileNetV2 have an accuracy of about 95\%. So even with precision scaling schemes, we can still retain effective malware detection capability.

%\vspace{-25em}

Table \ref{tab3} presents the comparison of the proposed technique with the existing malware detection techniques. We compare the performance of the proposed technique in terms of accuracy, F1 score, and recall. All the models in table \ref{tab3} focus on malware detection based on malware and benign features. Compared to the existing techniques the proposed PIM-based malware detection achieves high throughput without performance decay. It maintains a malware detection accuracy of 98\%. It is also evident that the proposed technique achieves efficient malware detection accuracy even in low-precision settings.

\begin{table}[htb!]
	\centering
	\caption{Comparison with existing HPC-based detection techniques}
	\vspace{-1em}
	\label{tab3}
	\scalebox{0.82}{
		\begin{tabular}{|c|c|c|c|c|c|}
			\hline
			
			\textbf{Model} &  \textbf{Accuracy} & \textbf{F1-score} & \textbf{Recall} \\%& \textbf{Latency}  & \textbf{Energy} \\
			& (\%) & &  \\%(s) &(J) \\
			\hline 
			OneR \cite{HPC_r1} & 0.81  & 0.81 & 0.82 \\ %& 1 & 1258\\
			\hline
			JRIP \cite{HPC_r1} & 0.83 & 0.83 & 0.84  \\% & 4 & 1504\\
			\hline
			PART \cite{HPC_r1}& 0.81 & 0.815 & 0.831 \\%& 6 & 2131\\
			\hline 
			J48 \cite{HPC_r1} & 0.82  & 0.82 & 0.82  \\%& 9 & 1801\\
			\hline
			Adaptive-HMD \cite{adaptive_hmd} & 0.853 & 0.853 & 0.858 \\%& 4 & 876 \\
			\hline
			SVM \cite{results_nn}& 0.739  & 0.736 & 0.772  \\%& - & -\\
			\hline
			RF \cite{results_nn} & 0.835 & 0.834 & 0.822 \\%& - & - \\
			\hline
			NN \cite{results_nn} & 0.811 & 0.811 & 0.816 \\%& - & - \\
			\hline
			SMO \cite{results_smo} & 0.932 & 0.933 & 0.931 \\%& 22  & 2466 \\
			\hline
			\textbf{Proposed} & \textbf{0.987} & \textbf{0.987} & \textbf{0.982} \\%& \textbf{10} & \textbf{1044} \\
			\hline

	\end{tabular}}
	%\vspace{-1.5em}
\end{table}

\subsection{PIM core and cluster characteristics}

The delay \& power for the PIM core and cluster are obtained from Synopsys Design Compiler using 28nm standard cell libraries from TSMC and are presented in Table \ref{table2}. The delay of a single 8-bit MAC performed within a cluster involves computations inside the PIM cores as well as communication among the cores. The power consumption of the cluster is that of all the cores and the core-to-core communication. The power and delay for intra and inter-subarray data transfers are obtained from \cite{lisa} and \cite{rowclone}. These metrics are used in the system-level performance evaluation of the PIM in the next subsections. 

\begin{table}[htb!]
	\centering
	\caption{Characteristics  of  Proposed PIM  components  in  28  nm node}
	\scriptsize
	\begin{tabular}[b]{p{3cm}|p{1.2cm}|p{1.2cm}|p{1.5cm}}
		
		\hline 
		Component & Delay (ns) & Power (mW) & Active Area ($\mu$m$^2$)  \\
		\hline 
		Proposed PIM Core & 0.8 & 2.7 & 4196.64 \\
		\hline 
		Proposed PIM Cluster (MAC Operation) & 6.4 & 8.2-11 & 37769.81 \\
		\hline
		Intra-Subarray Communication \cite{rowclone}* & 63.0 & 0.028 $\mu$J/comm & N/A \\
		\hline
		Inter-Subarray Communication \cite{lisa} for subarrays 1/7/15 hops away* & 148.5/ 196.5/ 260.5 & 0.09/ 0.12/ 0.17 $\mu$J/comm & N/A\\ 
		\hline 
		%NoC hop (Switch + wired Link)  & 1ns & 8188 & 1622 \\
		%    \hline
	\end{tabular}
	\label{table2}
	%\vspace{-0.5mm}
	\begin{footnotesize}
		\scriptsize\textit{*Represented in 28nm technology node}
	\end{footnotesize}
\end{table}

\subsection{Performance Evaluation}

\begin{figure}[h]
	\includegraphics[width=0.45\textwidth]{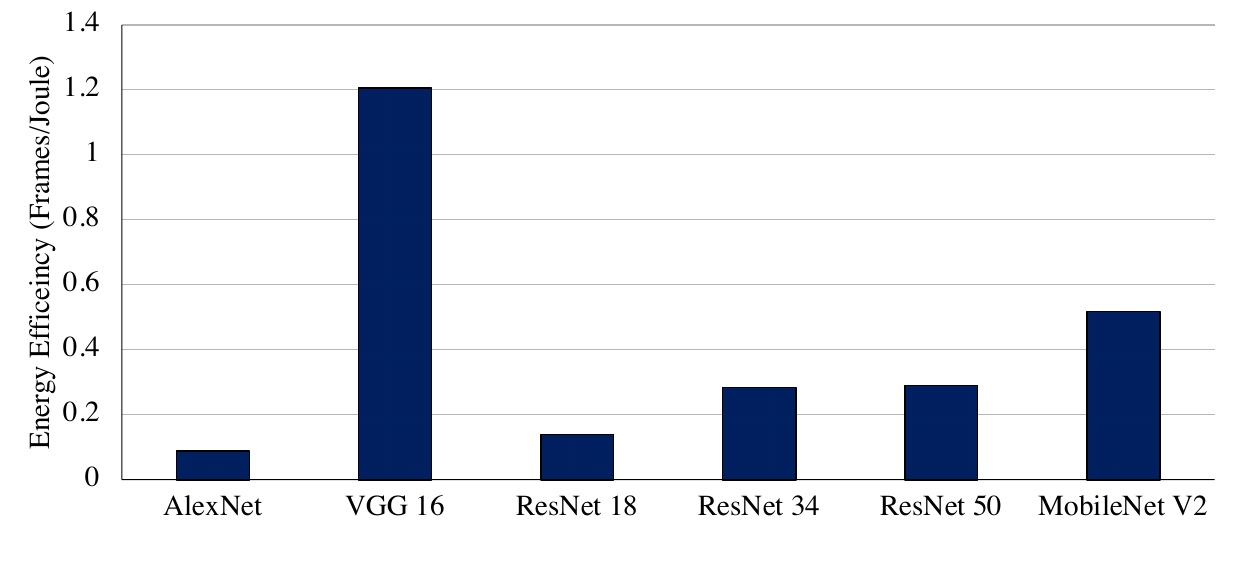}
	\caption{Comparison of Energy efficiency (Frames/Joules) for AlexNet, ResNet18, ResNet34, ResNet50, VGG16, and MobileNetV2 on the PIM accelerator}
	\label{fig:throughput}
\end{figure}

\begin{figure}[h]
	\includegraphics[width=0.45\textwidth]{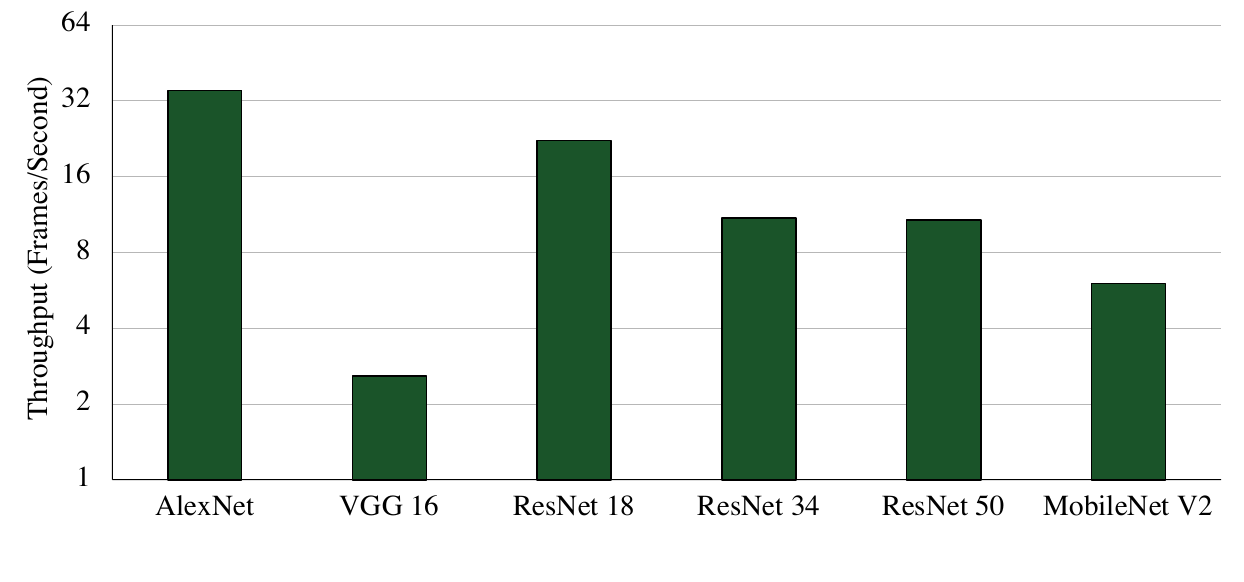}
	\caption{Comparison of Throughput (Frames/second) for AlexNet, ResNet18, ResNet34, ResNet50, VGG16, and MobileNetV2 on the PIM accelerator}
	\label{fig:energy}
\end{figure}

This section presents the comparative analysis of the algorithm implemented on PIM in terms of throughput (in Frames per second) and Energy Efficiency (Frames/Joules). For evaluation purposes, we have implemented AlexNet, ResNet 18, 34, 50, VGG16, and MobileNetV2 networks on the PIM accelerator.
Figure \ref{fig:throughput} presents comparisons of the throughput (in Frames per second) and  Figure \ref{fig:energy} energy efficiency (in Frames per Joule) of inference on all these CNNs deployed on the  PIM accelerator.

From Figure \ref{fig:throughput} it can be observed that the proposed PIM model is capable of performing malware detection tasks with an impressive performance of low latency. For example, ResNet-50, the largest network consisting of 50 layers and thirty-eight billion computations, is processed within 10 mS.

A similar trend is observed for energy consumption, from Figure \ref{fig:energy} it can be observed that the proposed PIM model is capable of performing malware detection tasks with high energy efficiency. This is because the PIM module supports 8-bit precision mode in order to perform the operations required for CNN acceleration. These tasks can be performed by distributing the data across the cluster which contains nine cores connected by a router which inherently offers a higher degree of parallelization and performs all the operations in comparatively fewer steps.

\subsection{Performance Comparison with State-of-the-Art Hardware Accelerators for CNN Implementation}

\begin{figure}[htb!]
	\centering
	\includegraphics[width=0.45\textwidth]{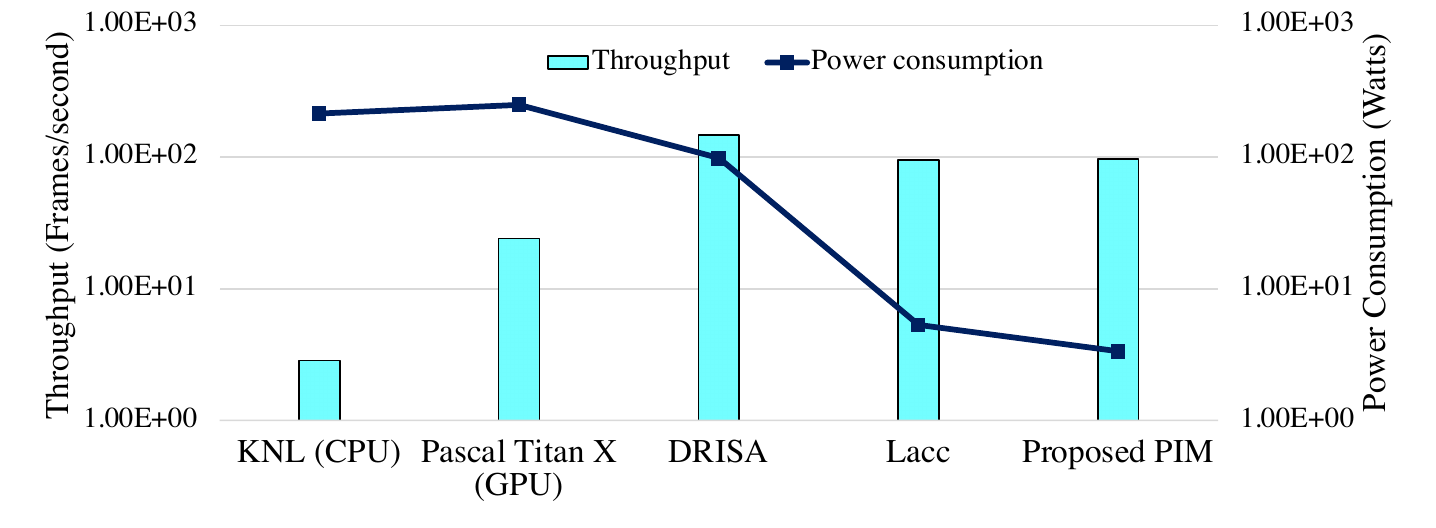}
	\caption{Comparative performance analysis of PIM with respect to state-of-the-art hardware accelerator architectures in terms of throughput (Frames/second) and power consumption (Watt)}
	\label{fig:compare_throughput}
\end{figure}

Performance is evaluated by comparing the proposed architecture with state-of-the-art hardware accelerator architectures in terms of power consumption (Watt) and throughput (Frames/second), as shown in Figure \ref{fig:compare_throughput}. 

As a proof of concept, we evaluate and implement AlexNet \cite{alex} on the proposed architecture with the 8-bit width precision. We envision a 256 PIM cluster arrangement in a DRAM chip as this configuration provides a fine balance between performance, power consumption, and on-chip area overhead. An input dimension of 224x224x3 has been considered for performance benchmarking with the other CNN accelerators and different operational modes of PIM. The PIM architectures under comparison in this section include DRAM-based bulk bit-wise processing devices DRISA \cite{DRISA}, and LUT-based PIM implemented on the DRAM platforms such as LAcc \cite{LACC}. On the other hand, the von Neumann devices under comparison are Intel Knights Landing (KNL), a state-of-the-art CPU \cite{knl}, and Pascal Titan X, a state-of-the-art GPU. 

It can be observed from Figure \ref{fig:compare_throughput}, that the PIM architectures in general outperform both the CPU and 
the GPU by a huge margin since all these PIMs can avoid the significant overhead and latency associated with off-chip communications, unlike the CPU and the GPU. In fact, the most computation-intensive 8-bit fixed-point operation mode PIM ideally provides 4.02$\times$, 45$\times$ higher throughput compared to Pascal Titan X GPU and Knights Landing Processor while being 74.62$\times$, 64.13$\times$ more energy-efficient. 

On the other hand, a relatively higher throughput is observed for DRISA \cite{DRISA} due to its ability to parallelize operations across multiple banks, albeit at significantly low power efficiency.
The benefits of adopting LUTs in order to utilize pre-calculated results instead of performing in-memory logic operations are convincingly demonstrated by LAcc \cite{LACC} which achieves impressive inference performance at quite a low power consumption.
From Figure \ref{fig:compare_throughput}, it is also observed that the proposed PIM outperforms DRISA and LAcc in both the throughput by 0.065$\times$, 1.09$\times$ as well as power efficiency by 29.25$\times$, 1.5$\times$ respectively for AlexNet inference.

%\vspace{-0.1em}

\section{Conclusion}
\label{conclusion}

In this paper, we proposed a PIM-based ML modeling technique for malware detection. The proposed approach not only achieves low latency for implementing malware detection task but also provides high energy efficiency. Which in turn makes the real-time malware detection task in embedded devices more approachable. The performance of the proposed PIM is evaluated by comparing it with state-of-the-art CPU, GPU, and other PIM architectures. The experimental results indicate that the proposed PIM is 74.62$\times$, 64.13$\times$ more energy-efficient and has 4.02$\times$, 45$\times$ higher throughput compared to the GPU and CPU respectively. It is also observed that the PIM is 1.5$\times$ energy efficient and has 1.09$\times$ higher throughput than other LUT-based PIM architecture. Multiple CNN models were trained using data that was precision scaled into 16-bit, 8-bit and 4-bit samples, to further aid the energy efficiency and throughput. The performance of these models is compared with state-of-the-art malware detectors. And from experimental results, it is evident that the proposed technique is efficient for malware detection as it doesn't experience any performance decay.

	\bibliographystyle{IEEEtran}
	\bibliography{reference.bib}
	
	% \appendix
	
\end{document}